\documentclass[runningheads]{llncs}
\usepackage{graphicx}
\usepackage{todonotes}
\usepackage{url}
\usepackage{paralist}
\usepackage{amsmath,amssymb} 
\usepackage{xcolor}
\usepackage{subfig}
\usepackage{hyperref}

\begin{document}
\title{On the relationship between calibrated predictors and unbiased volume estimation}
\titlerunning{Relationship between calibration and bias}
%
\author{Teodora Popordanoska \and     Jeroen Bertels \and
         Dirk Vandermeulen \and
         Frederik Maes \and
         Matthew B. Blaschko }
         

%
\authorrunning{T. Popordanoska et al.}
%
\institute{Center for Processing Speech and Images, Dept.\ ESAT, KU Leuven, Belgium
\email{teodora.popordanoska@kuleuven.be}}

%
\maketitle              
\begin{abstract}
Machine learning driven medical image segmentation has become standard in medical image analysis.  However, deep learning models are prone to overconfident predictions.  This has led to a renewed focus on \emph{calibrated predictions} in the medical imaging and broader machine learning communities.  Calibrated predictions are estimates of the probability of a label that correspond to the true expected value of the label conditioned on the confidence.  Such calibrated predictions have utility in a range of medical imaging applications, including surgical planning under uncertainty and active learning systems. 
At the same time it is often an accurate \emph{volume measurement} that is of real importance for many medical applications. This work investigates the relationship between model calibration and volume estimation. 
We demonstrate both mathematically and empirically that if the predictor is calibrated \textit{per image}, we can obtain the correct volume by taking an expectation of the probability scores per pixel/voxel of the image. Furthermore, we show that convex combinations of calibrated classifiers preserve volume estimation, but do not preserve calibration. Therefore, we conclude that having a calibrated predictor is a sufficient, but not necessary condition for obtaining an unbiased estimate of the volume. We validate  our theoretical findings empirically on a collection of 18 different (calibrated) training strategies on the tasks of glioma volume estimation on BraTS 2018, and ischemic stroke lesion volume estimation on ISLES 2018 datasets.

\end{abstract}

\section{Introduction}

In recent years the segmentation performance of CNNs improved dramatically. Despite these improvements, the adoption of automated segmentation systems into clinical routine is rather slow. Having calibrated model predictions would foster this relationship by providing the clinician with valuable information on failure detection or when manual intervention is needed~\cite{Jungo2020}. Reasoning under uncertainty is a central part of surgical planning~\cite{GILLMANN201826}, and uncertainty estimates are central to many active learning frameworks~\cite{wu2018active}. With this in mind, current research orients towards the design of automated segmentation systems that provide a measure of confidence alongside its predictions.

The empirical findings that modern CNNs are poorly calibrated~\cite{Guo2017} stimulated a variety of research to improve model calibration, including deep ensembles~\cite{Lakshminarayanan2017}, Bayesian NNs~\cite{Radford2012} and MC dropout~\cite{Gal2016}. The ability to deliver realistic segmentations would take calibration even further~\cite{Kohl2018,Baumgartner2019}. Remarkably, a large part of ongoing research stands orthogonal by design. Choosing a loss function that is consistent with a certain target metric \cite{Eelbode2020} often means the outputs cannot be interpreted as voxel-wise probabilities \cite{Bertels2020}, let alone that the predictions would be calibrated. Nonetheless, post-hoc calibration might come to the rescue and has proven to be competitive to training-time calibration \cite{Rousseau2021}.

While the automated segmentation is an important task in its own right, it is mostly used as an intermediary for calculating certain biomarkers. In that respect, volume is by far the most important biomarker in medical imaging. For example, tumor volume is a basic and specific response predictor in radiotherapy~\cite{Dubben1998} and the volume of an acute stroke lesion can be used to decide on the type of endovascular treatment~\cite{Goyal2016}. When the CNN outputs voxel-wise probabilities, they can be propagated to produce volumetric uncertainty~\cite{EatonRosen2018}. However, similar to its effect on calibration, the loss function determines how to calculate volume, and thus when the former is applicable~\cite{Bertels2020}.

This work bridges the gap between model calibration and volume estimation. There will be theoretical grounds that calibration error bounds volume error. This relationship is confirmed in an empirical validation on BraTS 2018~\cite{Menze2015,Bakas2017,Bakas2018} and ISLES 2018~\cite{Winzeck2018}. Furthermore, there is a clear empirical correlation between calibration error and volume error, and between object size and volume error. As a result, this work acknowledges and encourages research towards calibrated systems. Such systems not only provide additional robustness; they will also produce correct volume estimates.

\section{The relationship between calibration and volume bias}\label{sec:theory}

In this section, we demonstrate that calibrated uncertainty estimates are intimately related to unbiased volume estimates.  We develop novel mathematical results showing that calibration error upper bounds the absolute value of the volume bias (Proposition~\ref{prop:CEgeqBias}), which implies that as the calibration error goes to zero, the resulting function has unbiased volume estimates.  We further show that unbiased volume estimates do not imply that the classifier is calibrated, and that solely enforcing an unbiased classifier does not result in a calibrated classifier (Proposition~\ref{prop:zeroBiasNotImpliesZeroCE} and Corollary~\ref{prop:noMultBoundCEfromBias}).  This motivates our subsequent experimental study where we measure the empirical relationship between calibration error and volume bias in Sections~\ref{sec:Methods}.

\begin{definition}[Volume bias \cite{Bertels2020}]
Let $f$ be a function that predicts from an image/tomography $x$ the probability of each pixel/voxel $\{y_i\}_{i=1}^p$ belonging to a given class.  The \emph{volume bias} of $f$ is:
\begin{align}
    \operatorname{Bias}(f) := \mathbb{E}_{(x,y)\sim P}\left[  f(x) - y \right] .
    \label{def:bias}
\end{align}
\end{definition}

\begin{definition}[Calibration error \cite{naeini2015,kumar2019,wenger2020}]
\label{def:calibration-err}
The calibration error of $f : \mathcal{X} \to [0,1]$ is:
\begin{equation}
\operatorname{CE}(f) = \mathbb{E}_{(x,y)\sim P} \left[  \left| \mathbb{E}_{(x,y)\sim P} \left[ [y=1] \mid f(x) \right] - f(x) \right|  \right]
\end{equation}
\end{definition}
In plain English: A classifier is calibrated if its confidence score is equal to the probability of the prediction being correct.
Note that this definition is specific to a binary classification setting. The extension of binary calibration methods to multiple classes is usually done by reducing the problem of multiclass classification to $K$ one-vs.-all binary problems \cite{Guo2017}. 
The so-called \textit{marginal CE} \cite[Definition~2.4]{kumar2019}) is then measured as an average of the per-class CEs. 

\begin{proposition}\label{prop:CEgeqBias} 
The absolute value of dataset (respectively volume) bias is upper bounded by dataset (respectively volume) calibration error:
\begin{align}
    \operatorname{CE}(f) \geq |\operatorname{Bias}(f)|.
\end{align}
\end{proposition}
 \begin{proof}
    \begin{align}
    |\operatorname{Bias}(f)|
         =&\left| \mathbb{E}_{(x,y)\sim P}\left[  y - f(x) \right] \right|  \\
         =& \label{eq:BiasTwoTerms} \left| \underbrace{\mathbb{E}\left[ y - \mathbb{E} \left[ [y=1] \mid f(x) \right] \right]}_{=0} +  \mathbb{E}_{(x,y)\sim P}\left[  \mathbb{E} \left[ [y=1] \mid f(x) \right] - f(x) \right] \right| \\
           \leq & 
         \underbrace{\mathbb{E}_{(x,y)\sim P}\left[  \left| \mathbb{E} \left[ [y=1] \mid f(x) \right] - f(x) \right| \right]}_{=\operatorname{CE}(f)} .
         \label{eq:CalErrgeqEpsMinusSomething}
    \end{align}
    Focusing on the first term of the right hand side of \eqref{eq:BiasTwoTerms}, 
    \begin{align}
    \mathbb{E}\left[ y - \mathbb{E} [ [y=1] \mid f(x) \right] ] = \mathbb{E}[y] - \underbrace{\mathbb{E}[\mathbb{E} \left[ [y=1] \mid f(x) \right]]}_{=\mathbb{E}[y]}. \label{eq:EofYminusEofYgFx}
    \end{align}
    In the second term of the r.h.s., we may take the expectation with respect to $f(x)$ in place of $x$ as $f$ is a deterministic function.  This term is therefore also equal to $\mathbb{E}[y]$ by the law of total expectation.  
        Finally, the inequality in \eqref{eq:CalErrgeqEpsMinusSomething} is obtained due to the convexity of the absolute value and by application of Jensen's inequality.  
\qed
\end{proof}
We note that Proposition~\ref{prop:CEgeqBias} holds for all problem settings and class distributions. In multiclass settings, for each individual class we can obtain a bound by measuring the bias and the CE per class.

\begin{corollary}  
$\operatorname{CE}(f)=0$ implies that $f$ yields unbiased volume estimates.
\label{col:no_ece_no_bias}
\end{corollary}

\begin{proposition}\label{prop:zeroBiasNotImpliesZeroCE}
$\operatorname{Bias}(f)=0$ does not imply that $\operatorname{CE}(f)=0$.
\end{proposition}
\begin{proof}
 Consider the following example.
 Let the dataset consist of 100 positive and 200 negative points.
 Let $f_1$ and $f_2$ be binary classifiers that rank $1/4$ of the negative points with a score of zero, and the remaining negative points with a score of 0.25.
 Let $f_1$ rank the first half of the positive points with a score of one and the second half with a score of 0.25, and $f_2$ vice-versa. In this case, 
 $\operatorname{CE}(f_1)=0$ and $\operatorname{CE}(f_2)=0$, and therefore by Corollary~\ref{col:no_ece_no_bias}, $f_1$ and $f_2$ are both unbiased estimates of the volume.
 Let $f_3$ be a classifier that performs a convex combination (e.g. an average) of the scores of $f_1$ and $f_2$. 
 We note that a convex combination of unbiased estimators is unbiased \cite{lee1990u} and therefore $f_3$ is unbiased.
 Even though $f_3$ has a perfect accuracy and $\operatorname{Bias}(f_3)=0$, the scores are no longer calibrated, i.e., $\operatorname{CE}(f_3)\neq0$. \qed 
\end{proof}

\begin{corollary}\label{prop:noMultBoundCEfromBias}
There exists no multiplicative bound of the form $\operatorname{CE}(f) \leq \gamma |\operatorname{Bias}(f)|$ for some finite $\gamma>0$ (cf.\ \cite[Definition~2.2]{Eelbode2020}).
\end{corollary}
Thus we see that control over $\operatorname{CE}(f)$ minimizes an upper bound on $|\operatorname{Bias}(f)|$, but the converse is not true.  There are several implications of these theoretical results for the design of medical image analysis systems:
\begin{inparaenum}[(i)]
    \item Optimizing calibration error per-subject is an attractive method to simultaneously control the bias of the volume estimate, but we need to empirically validate if bias and CE are correlated in practice as we only know \emph{a priori} that one bounds the other; and
    \item optimizing volume bias alone (e.g.\ by empirical risk minimization) does not automatically give us the additional benefits of calibrated uncertainty estimates, and does not even provide a multiplicative bound on how poor the calibration error could be.
\end{inparaenum}
We consequently empirically evaluate the relationship between bias and calibration error in the remainder of this work.

\section{Empirical setup}\label{sec:Methods}

The empirical validation of our theoretical results will be performed by analyzing two segmentation tasks, each requiring a different distribution in the predicted confidences, with multiple different models, each trained with respect to a different loss function and subject to different post-hoc calibration strategies.

\textbf{Tasks}
The data from two publicly available medical datasets is used and two segmentation tasks are formulated as follows:
\begin{inparaenum}[(i)]
    \item Whole tumor segmentation using the BraTS 2018~\cite{Menze2015,Bakas2017,Bakas2018} (BR18) dataset. BR18 contains 285 multi-modal MR volumes with accompanying manual tumor delineations. Due to a rather low inter/intra-rater variability~\cite{Menze2015} the voxel-wise confidences will be distributed towards the high-confidence ranges;
    \item Ischemic core segmentation using the ISLES 2018~\cite{Winzeck2018} (IS18) dataset. IS18 contains data from 94 CT perfusion scans with manual delineations of the ischemic infarctions on co-registered DWI MR imaging. The identification of the ischemic infarction on CT perfusion data is generally considered non-trivial~\cite{Demeestere2017}. This means that a rather high intra/inter-rater variability is to be expected, which in turn will distribute the voxel-wise confidences across the entire range.
\end{inparaenum}
For both datasets there was a five-fold split of the data identical to~\cite{Bertels2020,Rousseau2021}.

\textbf{Models}
For the two former tasks the pre-trained and publicly available models from~\cite{Rousseau2021} are used. They investigated the effects of the loss function in combination with a multitude of different post-hoc calibration strategies on the Dice score and model calibration, but without any consideration to volume estimates or volume bias. Their base model shares a U-Net~\cite{Ronneberger2015} CNN architecture similar to~\cite{Isensee2018}. The three loss functions for the initial training were: 
\begin{inparaenum}[(i)]
    \item cross-entropy (CrE);
    \item soft-Dice (SD); and
    \item a combination of CrE pre-training with SD fine-tuning (CrE-SD). 
\end{inparaenum}    
In addition, these base models were calibrated using different post-hoc calibration strategies:
\begin{inparaenum}[(i)]
    \item Platt scaling and its variants (auxiliary network and fine-tuning); and
    \item two Monte Carlo (MC) dropout methods (MC-Dropout and MC-Center) with different positioning of the dropout layers.
\end{inparaenum} 
For further details on the exact training procedures the reader is referred to~\cite{Rousseau2021}. Nevertheless, it is important to note that the initial training and the post-hoc calibration was done on the training sets, and thus the predictions on the validation sets may be aggregated for further testing.

\textbf{Bias and ECE}
The Bias is calculated by direct implementation of Definition~\ref{def:bias}, i.e.\ the expectation of the probability scores per voxel. The CE from Definition~\ref{def:calibration-err} is a theoretical quantity that in practice is approximated by a binned estimator of the expected calibration error (ECE) (Equation~(3) from~\cite{Rousseau2021}). The ECE ranges from 0 to 1, with lower values representing better calibration. We use 20 bins for binning the CNN outputs. Following the example of \cite{Rousseau2021,Jungo2020}, we only consider voxels within the skull-stripped brain/lesion and report the mean per-volume Bias and ECE, as being more clinically relevant versions opposed to their dataset-level variants.

\textbf{Code}
The source code is available at~\url{https://github.com/tpopordanoska/calibration_and_bias}.

\section{Results and Discussion}
\begin{figure}[b!]
    \centering
    \includegraphics[width=\linewidth]{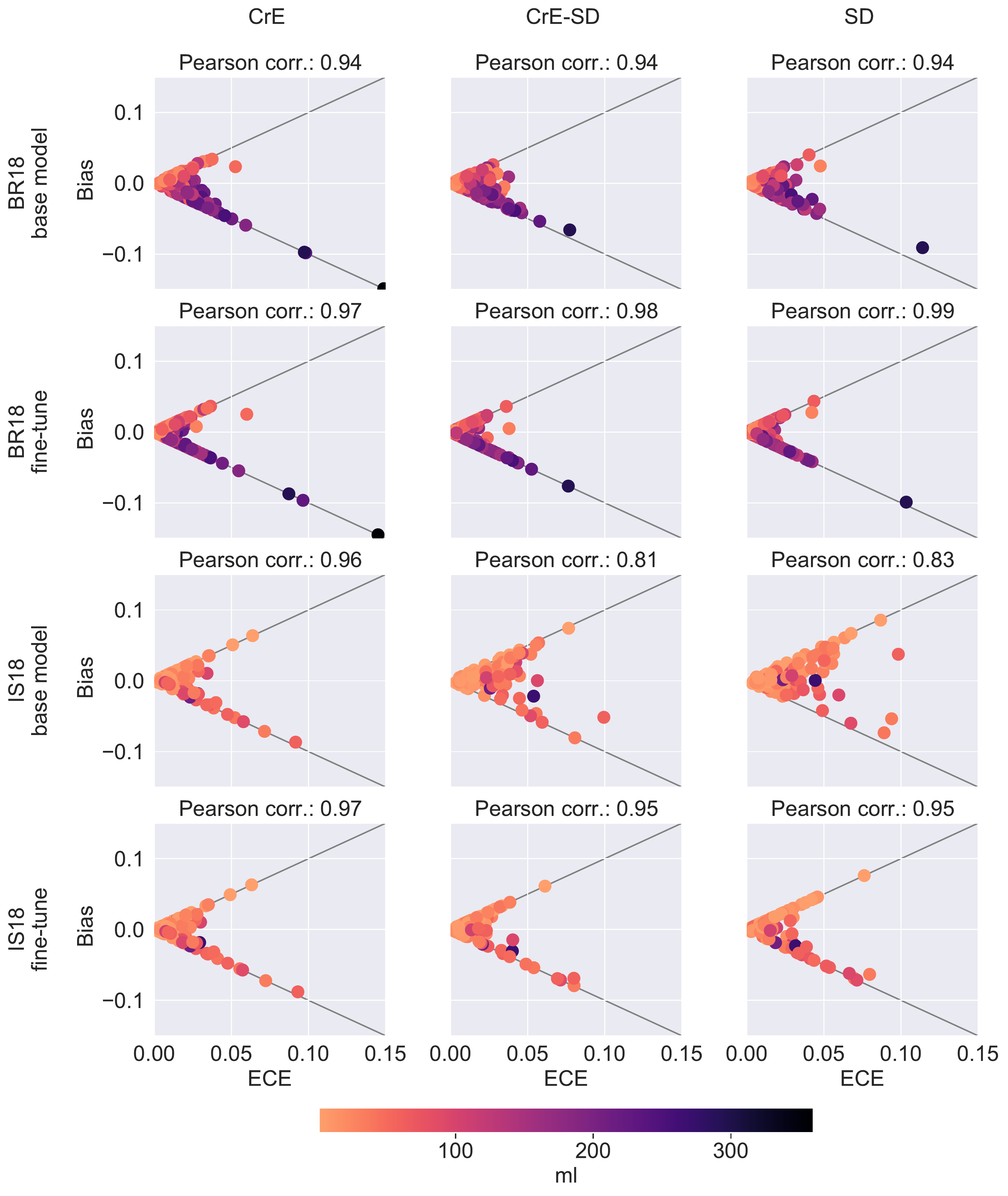}
    \caption{Scatter plots on BR18 (top two rows) and IS18 (bottom two rows), color-coded by tumor/lesion size (in ml). Every point in the plot represents an image. The Pearson correlation between per-volume ECE and absolute per-volume bias is shown above the plots. Note that the $\operatorname{Bias}$ is calculated in voxels, while the color-coding has the units converted to ml.}
    \label{fig:scatter_plots_per_volume}
\end{figure}

\begin{figure}[t]
    \centering
    \includegraphics[width=\linewidth]{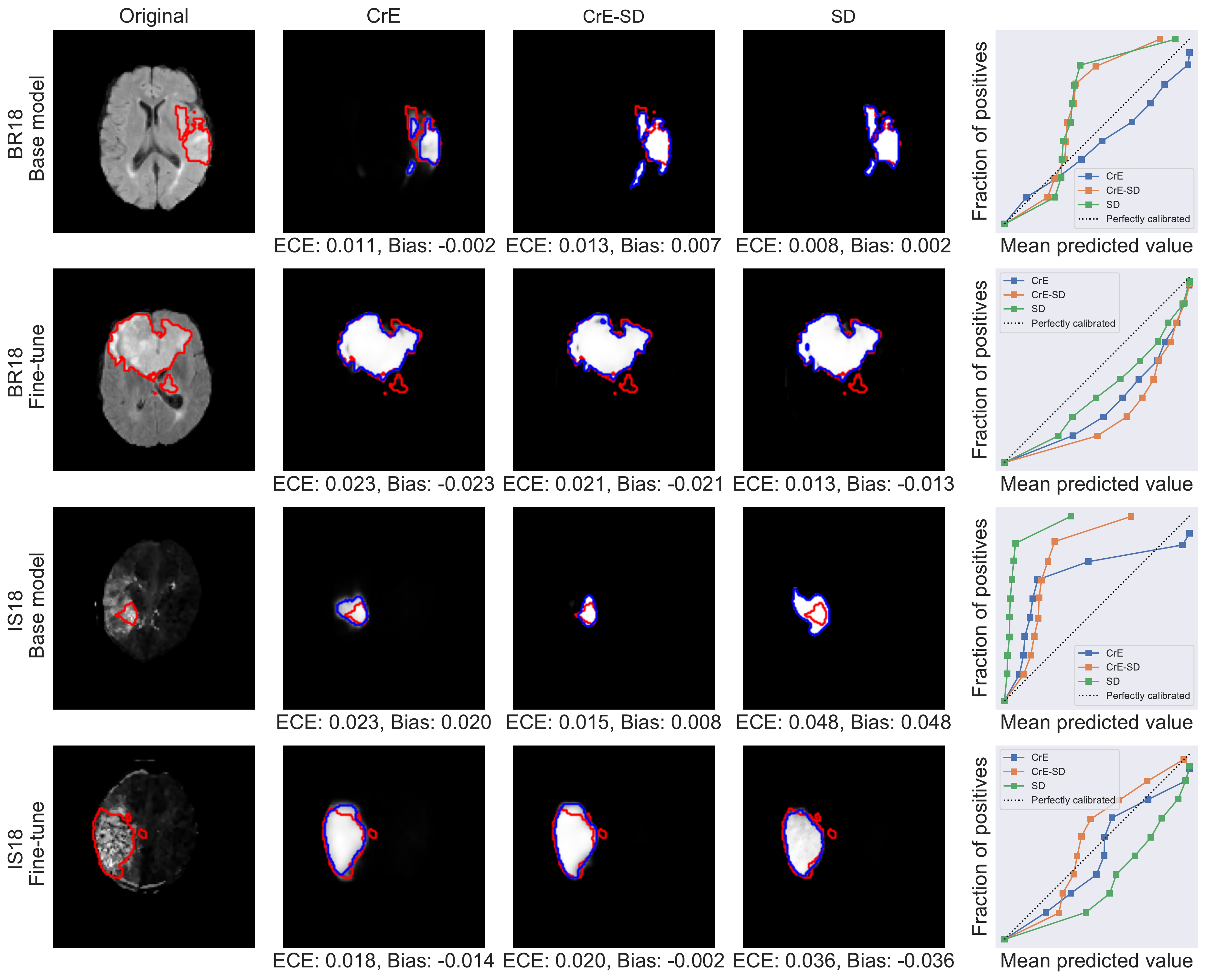}
    \caption{Qualitative examples of the predictions from a base model and a calibrated model with fine-tuning on BR18 (top two rows) and IS18 (bottom two rows). The red line represents the delineation of the ground truth. The predicted delineations after thresholding at 0.5 are overlayed in blue. The ECE and the bias (shown below the plot) are calculated for the selected slice. The last column shows the calibration curves for the volume. The images are chosen to be representative of different sizes of tumors/lesions and the middle slice of the volume is shown.}
    \label{fig:examples}
\end{figure}

Fig.~\ref{fig:scatter_plots_per_volume} shows scatter plots of ECE and $\operatorname{Bias}$ for the base model and a calibrated model with the fine-tuning strategy for BR18 and IS18. 
We can confirm that, as predicted by Proposition~\ref{prop:CEgeqBias}, all the points (volumes measured in ml) lie in-between the lines ECE = $\pm \operatorname{Bias}$.
We note further that the correlation between ECE and $\operatorname{Bias}$ is strictly higher for a calibrated model compared to the base model. There are many volumes for which $\operatorname{ECE} =  |\operatorname{Bias}|$ holds, which supports the findings in~\cite{Jungo2020} where per-volume calibration tends to be off, either resulting in a complete under- or over-estimation. This is also visible in Fig.~\ref{fig:examples}, when the calibration curve lies below or above the unity line $\operatorname{ECE} =  |\operatorname{Bias}|$.

\begin{figure}[t]
    \centering
    \includegraphics[width=\linewidth]{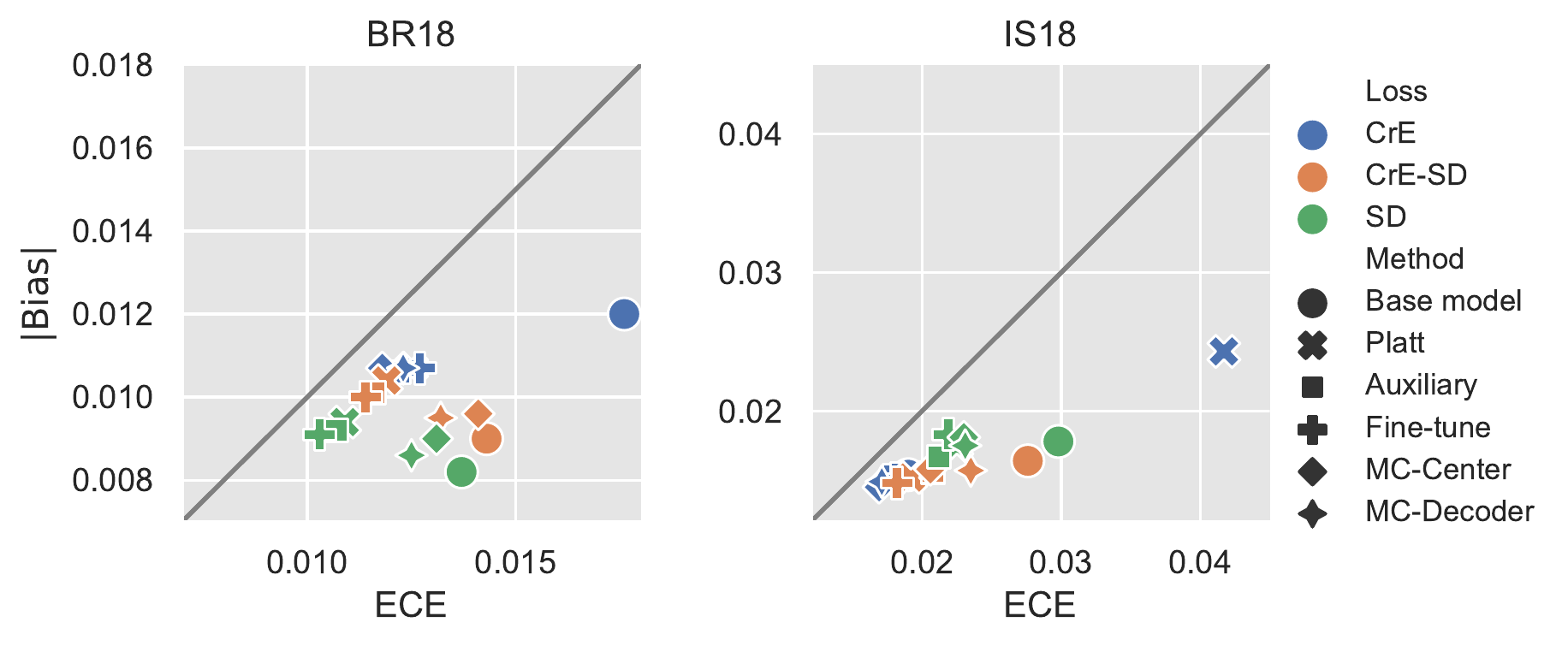}
    \caption{Scatter plots of ECE versus $|\operatorname{Bias}|$ for all combinations of loss functions and calibration methods. The Pearson correlation between $|\operatorname{Bias}|$ (computed as mean of absolute per-volume biases) and ECE (mean per-volume ECE) is  0.30 $\pm$ 0.21 for BR18 and 0.91 $\pm$ 0.04 for IS18.}
    \label{fig:scatter_all_methods}
\end{figure}

The ECE and $\operatorname{Bias}$ for all 18 models are visually presented in a scatter plot in Fig.~\ref{fig:scatter_all_methods}. Analogously to \cite{Rousseau2021}, we find that for the BR18 data only the models trained with SD are on the Pareto front. However, contrary to their result that MC methods are Pareto dominated if optimizing the Dice score is of interest, we observe that the MC-Decoder calibration strategy is Pareto-efficient for the clinical application of measuring volume.  For IS18, the model trained with CrE and calibrated with the MC strategy Pareto dominates all the rest. 
Therefore, we conclude that when selecting a model for volume estimation, prioritization of lowering the ECE over choosing the right training loss is advised. This somewhat refines the findings in~\cite{Bertels2020} where CrE outperformed SD variants in terms of volume bias on a dataset level.
This is further exemplified in Fig.~\ref{fig:examples} which shows visually on selected slices of different-sized volumes that the ECE is a more reliable predictor of the $\operatorname{Bias}$ than the loss function used during training.  

As an additional experiment on the difference between datasets with different levels of inter-rater variability, we calculated correlations between the mean absolute per-volume  $\operatorname{Bias}$ and mean per-volume ECE for the 18 settings on the BR18 dataset separated into high grade (HGG) and low grade glioma (LGG). The Pearson correlation is 0.15 $\pm$ 0.23 for HGG and 0.39 $\pm$ 0.20 for LGG. Compared to the analogous result in the caption of Figure~\ref{fig:scatter_all_methods}, we observe again that the correlation is higher for the data with higher uncertainty (LGG and IS18).

Table~\ref{tab:corr_coeff_table} shows the Pearson correlation coefficients with error bars \cite{Bowley1928} between the per-volume bias and the tumor/lesion size 
for both datasets. Negative correlations are also visible in the color-coded volumes in Fig.~\ref{fig:scatter_plots_per_volume}. All models have a tendency to underestimate large volumes (negative bias) and overestimate small volumes (positive bias). This trend was also observed in~\cite{Bertels2020,Tilborghs2020} where simple post-hoc training-set regression was used for recalibration.
We further observe that the correlation is stronger for the volumes in the BR18 than IS18 data. However, there is no consistent finding that holds for both datasets (e.g. the effect on the correlation is not influenced by the loss or post-calibration method). 
Additionally, we calculated the Spearman $\rho$ and Kendall $\tau$ coefficients for all settings. In all cases there are significant non-zero correlations and the median values for BR18 are -0.43 ($\rho$) and -0.29 ($\tau$), and for IS18 -0.43 ($\rho$) and -0.30 ($\tau$).

\begingroup
\setlength{\tabcolsep}{3pt} 
\renewcommand{\arraystretch}{1.4} 
\begin{table}[h]
\caption{Pearson correlation coefficients between per-volume bias and volume size for BR18 and IS18.}
\label{tab:corr_coeff_table}
\centering
\scriptsize
    \begin{tabular}{c|ccc|ccc}
    \textit{loss\(\rightarrow\)} & CrE  & CrE-SD     & SD &   CrE  & CrE-SD     & SD   \\ \hline
    method $\downarrow$      &       &   BR18    &   &  &  IS18   \\ \hline
    base model  & -0.73  $\pm$ 0.03 & -0.53  $\pm$ 0.04 & -0.48  $\pm$ 0.05 & -0.39  $\pm$ 0.09 & -0.23  $\pm$ 0.10 & -0.20  $\pm$ 0.10 \\ 
    Platt       & -0.58  $\pm$ 0.04 & -0.67  $\pm$ 0.03 & -0.62  $\pm$ 0.04 & -0.50  $\pm$ 0.08 & -0.39  $\pm$ 0.09 & -0.48  $\pm$ 0.08 \\ 
    auxiliary   & -0.57  $\pm$ 0.04 & -0.65  $\pm$ 0.03 & -0.61  $\pm$ 0.04 & -0.32  $\pm$ 0.09 & -0.31  $\pm$ 0.09 & -0.43  $\pm$ 0.08 \\ 
    fine-tune   & -0.59  $\pm$ 0.04 & -0.62  $\pm$ 0.04 & -0.55  $\pm$ 0.04 & -0.37  $\pm$ 0.09 & -0.36  $\pm$ 0.09 & -0.49  $\pm$ 0.08 \\ 
    MC-Decoder  & -0.55  $\pm$ 0.04 & -0.47  $\pm$ 0.05 & -0.42  $\pm$ 0.05 & -0.23  $\pm$ 0.10 & -0.27  $\pm$ 0.10 & -0.36  $\pm$ 0.09 \\ 
    MC-Center   & -0.53  $\pm$ 0.04 & -0.49  $\pm$ 0.04 & -0.49  $\pm$ 0.04 & -0.27  $\pm$ 0.10 & -0.24  $\pm$ 0.10 & -0.39  $\pm$ 0.09 \\ 
    \hline
    \end{tabular}
\end{table}
\endgroup

\noindent
\textbf{Limitations}
We showed theoretically and empirically that the CE is an upper bound for the $\operatorname{Bias}$, and that the $\operatorname{Bias}$ does not provide a bound on CE. Nonetheless, we did not properly characterize when CE may be strictly larger than this quantity and whether there is a meaningful interpretation of $\operatorname{CE}-|\operatorname{Bias}|$. Furthermore, we proved that a convex combination of calibrated classifiers does not preserve calibration (although it preserves unbiasedness), but we did not explore ways to combine calibrated models to obtain a calibrated predictor.

Designing a calibrated estimator is a non-trivial task and the field of confidence calibration is an active area of research. 
However, investing resources to find out which calibration strategy works best for the problem at hand is of high importance, especially in medical applications.

Finally, we wish to emphasize the distinction between dataset and per-volume ECE. Often in clinical settings, the information about per-subject calibration is of higher importance. A zero dataset level ECE implies zero dataset bias, however, the subject-level biases may very well be non-zero.  This subject is treated less frequently in the calibration literature.

\section{Conclusions}
In this work, it was shown that the importance of confidence calibration goes beyond using the predicted voxel-wise confidences for clinical guidance and robustness. More specifically, there is theoretical and empirical evidence that calibration error bounds the error of volume estimation, which still is one of the most relevant biomarkers calculated further downstream. Since the converse relationship does not hold, the direct optimization of calibration error is to be preferred over the optimization of volume bias alone.

\section*{Acknowledgments}
 This research received funding from the Flemish Government under  the  ``Onderzoeksprogramma  Artifici\"{e}le  Intelligentie (AI) Vlaanderen'' programme.  
J.B.\ is part of NEXIS (\url{www.nexis-project.eu}), a project that has received funding from the European Union's Horizon 2020 Research and Innovations Programme (Grant Agreement \#780026).

\bibliographystyle{splncs04}
\bibliography{refs}

\end{document}


\title{On the relationship between calibrated predictors and unbiased volume estimation: Supplementary material}
\titlerunning{Relationship between calibration and bias}
%
\author{Teodora Popordanoska \and     Jeroen Bertels \and
         Dirk Vandermeulen \and
         Frederik Maes \and
         Matthew B. Blaschko }
%
\authorrunning{T. Popordanoska et al.}
%
\institute{Center for Processing Speech and Images, Dept.\ ESAT, KU Leuven, Belgium
\email{teodora.popordanoska@kuleuven.be}}

%
\maketitle              
\begin{figure}[ht]
    \centering
    \includegraphics[width=0.8\linewidth]{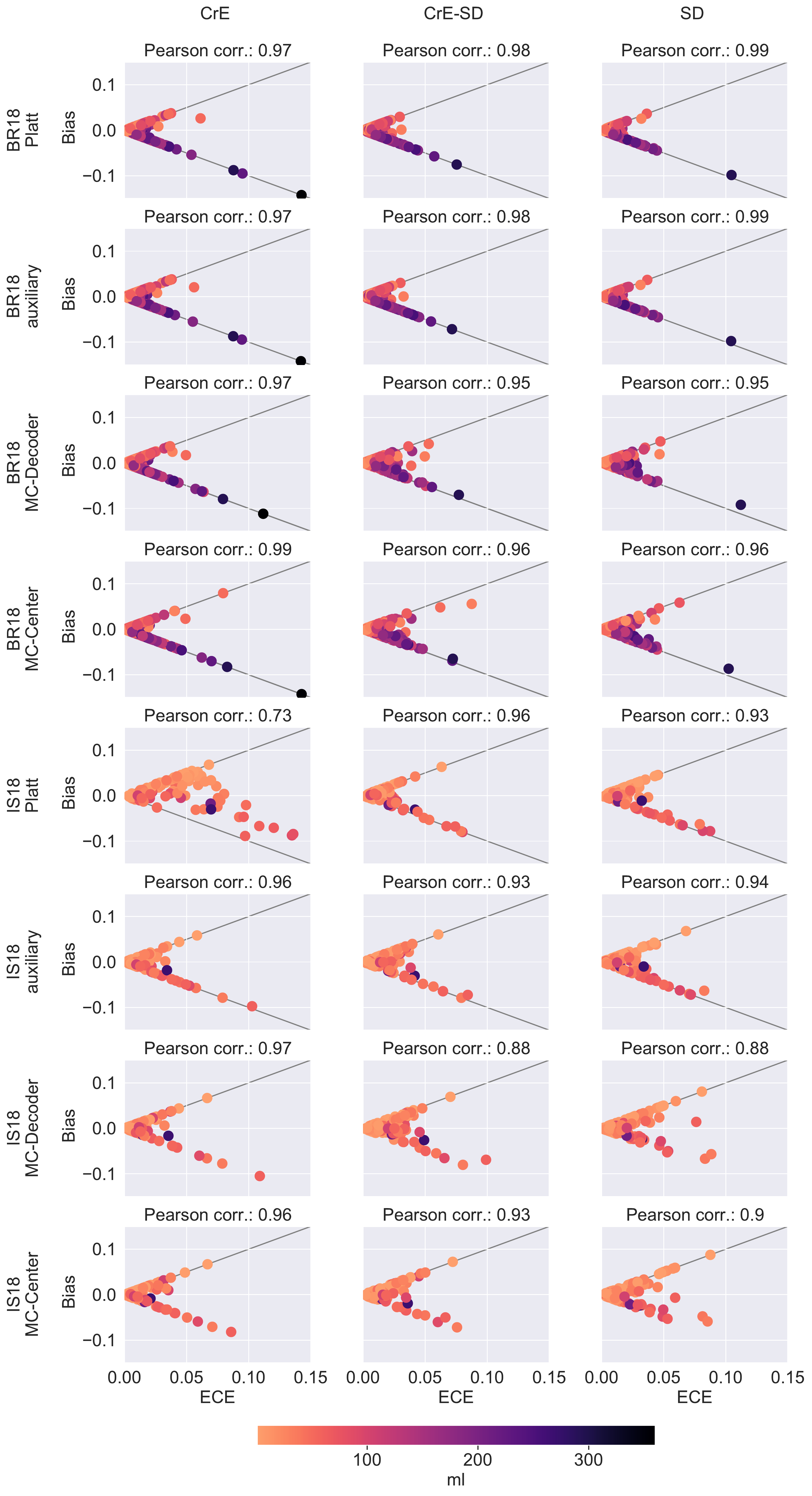}
    \caption{Scatter plots on BR18 (top four rows) and IS18 (bottom four rows), color-coded by tumor/lesion size (in ml) for the remaining four calibration strategies. Every point in the plot represents an image. The Pearson correlation between per-volume ECE and absolute per-volume bias is shown above the plots.}
    \label{fig:my_label}
\end{figure}

\begingroup
\renewcommand{\arraystretch}{1.5} 
\begin{table}
\caption{Summary of mean per-volume ECE and mean absolute per-volume Bias for all combinations of loss functions and methods on BR18.}
\centering
\tiny
\label{tab:brats-table}
    \begin{tabular}{c|ccc|ccc}
    \textit{loss\(\rightarrow\)} & CrE  & CrE-SD     & SD &   CrE  & CrE-SD     & SD   \\ \hline
    method $\downarrow$      &       &   $|\operatorname{Bias}|$    &   &  &  ECE   \\ \hline
    base model  & 0.0120  $\pm$ .0009 & 0.0090  $\pm$ .0008 & 0.0082  $\pm$ .0008 & 0.0176  $\pm$ .0008 & 0.0143  $\pm$ .0008 & 0.0137  $\pm$ .0008 \\ 
    Platt       & 0.0107  $\pm$ .0008 & 0.0104  $\pm$ .0008 & 0.0094  $\pm$ .0008 & 0.0125  $\pm$ .0008 & 0.0119  $\pm$ .0008 & 0.0109  $\pm$ .0008 \\ 
    auxiliary   & 0.0106  $\pm$ .0008 & 0.0101  $\pm$ .0008 & 0.0092  $\pm$ .0008 & 0.0122  $\pm$ .0008 & 0.0116  $\pm$ .0008 & 0.0107  $\pm$ .0008 \\ 
    fine-tune   & 0.0107  $\pm$ .0008 & 0.0100  $\pm$ .0008 & 0.0091  $\pm$ .0008 & 0.0127  $\pm$ .0008 & 0.0114  $\pm$ .0008 & 0.0103  $\pm$ .0008 \\ 
    MC-Decoder  & 0.0107  $\pm$ .0007 & 0.0095  $\pm$ .0008 & 0.0086  $\pm$ .0008 & 0.0123  $\pm$ .0007 & 0.0132  $\pm$ .0009 & 0.0125  $\pm$ .0008 \\ 
    MC-Center   & 0.0107  $\pm$ .0008 & 0.0096  $\pm$ .0009 & 0.0090  $\pm$ .0008 & 0.0118  $\pm$ .0008 & 0.0141  $\pm$ .0009 & 0.0131  $\pm$ .0009 \\ 
    \hline
    \end{tabular}
\end{table}

\begin{table}
\caption{Summary of mean per-volume ECE and mean absolute per-volume Bias for all combinations of loss functions and methods on IS18.}
\label{tab:isles-table}
\centering
\tiny
    \begin{tabular}{c|ccc|ccc}
    \textit{loss\(\rightarrow\)} & CrE  & CrE-SD     & SD &   CrE  & CrE-SD     & SD   \\ \hline
    method $\downarrow$      &       &   $|\operatorname{Bias}|$    &   &  &  ECE   \\ \hline
    base model  & 0.0154  $\pm$ .0017 & 0.0164  $\pm$ .0018 & 0.0178  $\pm$ .0019 & 0.0190  $\pm$ .0016 & 0.0276  $\pm$ .0019 & 0.0298  $\pm$ .0022 \\ 
    Platt       & 0.0243  $\pm$ .0022 & 0.0152  $\pm$ .0018 & 0.0175  $\pm$ .0019 & 0.0417  $\pm$ .0031 & 0.0192  $\pm$ .0017 & 0.0225  $\pm$ .0018 \\ 
    auxiliary   & 0.0153  $\pm$ .0017 & 0.0156  $\pm$ .0018 & 0.0167  $\pm$ .0018 & 0.0180  $\pm$ .0017 & 0.0208  $\pm$ .0017 & 0.0212  $\pm$ .0018 \\ 
    fine-tune   & 0.0151  $\pm$ .0017 & 0.0148  $\pm$ .0018 & 0.0183  $\pm$ .0018 & 0.0181  $\pm$ .0017 & 0.0182  $\pm$ .0018 & 0.0219  $\pm$ .0018 \\ 
    MC-Decoder  & 0.0149  $\pm$ .0018 & 0.0157  $\pm$ .0018 & 0.0175  $\pm$ .0017 & 0.0171  $\pm$ .0018 & 0.0235  $\pm$ .0018 & 0.0231  $\pm$ .0020 \\ 
    MC-Center   & 0.0145  $\pm$ .0016 & 0.0158  $\pm$ .0017 & 0.0181  $\pm$ .0018 & 0.0169  $\pm$ .0016 & 0.0206  $\pm$ .0016 & 0.0230  $\pm$ .0019 \\ 
    \hline
    \end{tabular}
\end{table}

\endgroup

\begin{table}
\caption{Tabulated list of parameters used in various calibration strategies.}
\label{tab:training-parameters-table}
\centering
\scriptsize
    \begin{tabular}{c|c|c|c}
        \textit{parameter\(\rightarrow\)} & batch size & initial learning rate & epochs \\ 
        \hline
        base model  &  2 for BR18; 4 for IS18 & $10^{-3}$ & until convergence \\ 
        Platt       &  64 z-slices & $ 5\cdot 10^{-3}$ & max 50 \\ 
        auxiliary   &  64 z-slices & $ 5\cdot 10^{-3}$ & max 50 \\ 
        fine-tune   &  2 3D volumes & $10^{-3}$ for SD; $10^{-4}$ for CrE & max 50\\ 
        MC methods  &  2 3D volumes & best of $\{10^{-3}, 10^{-4}, 10^{-5}\}$ & max 50 \\ 
        \hline
    \end{tabular}
\end{table}